\documentclass[conference]{IEEEtran}
\IEEEoverridecommandlockouts
\usepackage{cite}
\usepackage{amsmath,amssymb,amsfonts}
\usepackage{algorithmic}
\usepackage{graphicx}
\usepackage{textcomp}
\usepackage{xcolor}
\ifCLASSOPTIONcompsoc
  \usepackage[caption=false,font=normalsize,labelfont=sf,textfont=sf]{subfig}
\else
  \usepackage[caption=false,font=footnotesize]{subfig}
\fi

\DeclareMathOperator\llr{LLR}
\usepackage{mathtools}
\usepackage{bm}
\usepackage{booktabs}
\usepackage{multirow}
\usepackage[per-mode=symbol]{siunitx}
\usepackage{pgfplots}
\usetikzlibrary{matrix}
\pgfplotsset{compat=newest}
\usetikzlibrary{external}
\usepackage[acronym]{glossaries}
\newacronym{AED}{AED}{Automorphism Ensemble Decoding}
\newacronym{AWGN}{AWGN}{Additive White Gaussian Noise}
\newacronym{ASIC}{ASIC}{Application-Specific Integrated Circuit}
\newacronym{BER}{BER}{Bit Error Rate}
\newacronym{BP}{BP}{Belief Propagation}
\newacronym{BPSK}{BPSK}{Binary Phase Shift Keying}
\newacronym{BSMC}{BSMC}{Binary Symmetric Memoryless Channel}
\newacronym{CRC}{CRC}{Cyclic Redundancy Check}
\newacronym{DRAM}{DRAM}{Dynamic Random Access Memory}
\newacronym{eMBB}{eMBB}{enhanced mobile broadband}
\newacronym{ExPOS}{ExPOS}{Extended Threshold Partial Order}
\newacronym{FD-SOI}{FD-SOI}{Fully Depleted Silicon on Insulator}
\newacronym{FEC}{FEC}{Forward Error Correction}
\newacronym{FER}{FER}{Frame Error Rate}
\newacronym{GA}{GA}{Gaussian Approximation}
\newacronym{GSM}{GSM}{Global System for Mobile Communications}
\newacronym{IP}{IP}{Intellectual Property}
\newacronym{IoT}{IoT}{Internet of Things}
\newacronym{LDPC}{LDPC}{Low Density Parity Check}
\newacronym{LLR}{LLR}{Log-Likelihood Ratio}
\newacronym{LTE-A}{LTE-A}{Long Term Evolution Advanced}
\newacronym{MAP}{MAP}{Maximum a Posteriori}
\newacronym{ML}{ML}{Maximum Likelihood}
\newacronym{MS}{MS}{Min-Sum}
\newacronym{NR}{NR}{New Radio}
\newacronym{PAR}{PAR}{Placement and Routing}
\newacronym{PFT}{PFT}{Polar Factor Tree}
\newacronym{PVT}{PVT}{Process, Voltage and Temperature}
\newacronym{QC}{QC}{quasi-cyclic}
\newacronym{RM}{RM}{Reed-Muller}
\newacronym{SC}{SC}{Successive Cancellation}
\newacronym{SCAL}{SCAL}{Successive Cancellation Automorphism List}
\newacronym{SCAN}{SCAN}{Soft Cancellation}
\newacronym{SCL}{SCL}{Successive Cancellation List}
\newacronym{SRAM}{SRAM}{Static Random Access Memory}
\newacronym{SNR}{SNR}{Signal-to-Noise-Ratio}
\newacronym{SOA}{SOA}{State-of-the-Art}
\newacronym{SoC}{SoC}{System-on-Chip}
\newacronym{TDP}{TDP}{Thermal design power}
\newacronym{UMTS}{UMTS}{Universal Mobile Telecommunications System}
\newacronym{URLLC}{URLLC}{Ultra-Reliable Low Latency Communications}
\newacronym{PM}{PM}{Path Metric}
\newacronym{R0}{R0}{Rate-0}
\newacronym{REP}{REP}{Repetition}
\newacronym{SPC}{SPC}{Single Parity-Check}
\newacronym{HDD}{HDD}{Hard Decision Decoding}
\newacronym{BLTA}{BLTA}{Block Lower-Triangular Affine}
\newacronym[firstplural=Equivanlence Classes (ECs)]{EC}{EC}{Equivalence Class}
\glssetnoexpandfield{first}
\glssetnoexpandfield{firstpl}
\newglossaryentry{SSC}{
    type={acronym}, 
    sort={Simplified Successive-Cancellation},  
    name={SSC}, 
    short={SSC}, 
    long={Simplified Successive-Cancellation},
    first={\ifglsused{SSC}{SSC}{\glsunset{SC}Simplified Successive-Cancellation (SSC)}}, 
    description={Simplified Successive-Cancellation}
}
\glssetnoexpandfield{first}
\glssetnoexpandfield{firstpl}
\newglossaryentry{FSSC}{
    type={acronym}, 
    sort={Fast Simplified Successive-Cancellation},  
    name={FSSC}, 
    short={Fast-SSC}, 
    long={Fast Simplified Successive-Cancellation},
    first={\ifglsused{SSC}{Fast-SSC}
        {\ifglsused{SC}{Fast Simplified SC (Fast-SSC)}{\glsunset{SC}Fast Simplified Successive-Cancellation (Fast-SSC)}}}, 
    description={Fast Simplified Successive-Cancellation}
}
\glssetnoexpandfield{first}
\glssetnoexpandfield{firstpl}
\newglossaryentry{SSCL}{
    type={acronym}, 
    sort={Simplified Successive-Cancellation List},  
    name={SSCL}, 
    short={SSCL}, 
    long={Simplified Successive-Cancellation List},
    first={\ifglsused{SCL}{Simplified SCL (SSCL)}{\glsunset{SCL}Simplified Successive-Cancellation List (SSCL)}}, 
    description={Simplified Successive-Cancellation List}
}
\glssetnoexpandfield{first}
\glssetnoexpandfield{firstpl}
\newglossaryentry{FSSCL}{
    type={acronym}, 
    sort={Fast Simplified Successive-Cancellation List},  
    name={Fast-SSCL}, 
    short={Fast-SSCL}, 
    long={Fast Simplified Successive-Cancellation List},
    first={\ifglsused{SSCL}{Fast-SSCL}{\glsunset{SSCL}Fast Simplified Successive-Cancellation List (Fast-SSCL)}}, 
    description={Fast Simplified Successive-Cancellation List}
}
\glssetnoexpandfield{first}
\glssetnoexpandfield{firstpl}
\newglossaryentry{SSCAL}{
    type={acronym}, 
    sort={Simplified Successive-Cancellation Automorphism List},  
    name={SSCAL}, 
    short={SSCAL}, 
    long={Simplified Successive-Cancellation Automorphism List},
    first={\ifglsused{SCAL}{Simplified SCAL (SSCAL)}{\glsunset{SCAL}Simplified Successive-Cancellation Automorphism List (SSCAL)}}, 
    description={Simplified Successive-Cancellation Automorphism List}
}

\DeclareSIUnit[]{\bit}{bit}

\newcommand\mmsq[1]{\SI{#1}{\milli\meter\squared}}

\newcommand{\ie}{i.\,e.,\ }
\newcommand{\eg}{e.\,g.,\ }

\newcommand\dB[1]{\SI{#1}{\decibel}}
\newcommand\nm[1]{\SI{#1}{\nano\meter}}

\newcommand{\AED}{\gls{AED}\ }

\newcommand{\FER}{\gls{FER}\ }

\newcommand{\LLR}{\gls{LLR}\ }
\newcommand{\ML}{\gls{ML}\ }

\newcommand{\PMs}{\glspl{PM}\ }

\newcommand{\SCAL}{\gls{SCAL}\ }
\newcommand{\SCL}{\gls{SCL}\ }
\newcommand{\SC}{\gls{SC}\ }
\newcommand{\SNR}{\gls{SNR}\ }
\newcommand{\SPC}{\gls{SPC}\ }

\usepackage{balance}
\usepackage{blindtext}
\usepackage[shortcuts]{extdash}

\begin{document}

\title{
    Successive Cancellation Automorphism List Decoding of Polar Codes
    \thanks{
        The authors acknowledge the financial support by the Federal Ministry of
        Education and Research of Germany in the project “Open6GHub” (grant
        numbers: 16KISK019 and 16KISK004).
    }
}

\author{
    \IEEEauthorblockN{
        Lucas Johannsen\IEEEauthorrefmark{1}\IEEEauthorrefmark{2},
        Claus Kestel\IEEEauthorrefmark{1},
        Marvin Geiselhart\IEEEauthorrefmark{3},
        Timo Vogt\IEEEauthorrefmark{2},
        Stephan ten Brink\IEEEauthorrefmark{3} and
        Norbert Wehn\IEEEauthorrefmark{1}
    }
    \IEEEauthorblockA{
        \IEEEauthorrefmark{1}
        RPTU Kaiserslautern-Landau, 67663 Kaiserslautern, Germany,
    }
    \IEEEauthorblockA{
        \IEEEauthorrefmark{2}
        Koblenz University of Applied Sciences, 56075 Koblenz, Germany,
    }
    \IEEEauthorblockA{
        \IEEEauthorrefmark{3}
        Institute of Telecommunications, 
        University of Stuttgart, 70569 Stuttgart, Germany \\
        Email: \{lucas.johannsen, kestel, norbert.wehn\}@rptu.de,
        \{johannsen, vogt\}@hs-koblenz.de,\\
        \{geiselhart,tenbrink\}@inue.uni-stuttgart.de
    }
}

\maketitle

\begin{abstract}
    The discovery of suitable automorphisms of polar codes gained a lot of attention
by applying them in \gls{AED} to improve the error-correction performance, 
especially for short block lengths. This paper introduces \SCAL decoding of 
polar codes as a novel application of automorphisms in advanced \SCL decoding. 
Initialized with L permutations sampled from the automorphism group, a 
superposition of different noise realizations and path splitting takes place 
inside the decoder. In this way, the \SCAL decoder automatically adapts to the 
channel conditions and outperforms the error-correction performance 
of conventional \SCL decoding and \gls{AED}. For a polar code of length 128, 
\SCAL performs near \gls{ML} decoding with L=8, in contrast to M=16 needed 
decoder cores in \gls{AED}. \gls{ASIC} implementations in a \nm{12} 
technology show that high-throughput, pipelined \SCAL decoders 
outperform \AED in terms of energy efficiency and power density, and 
\SCL decoders additionally in area efficiency.

\end{abstract}
\glsresetall

\begin{IEEEkeywords}
Polar Code, Automorphisms, Successive Cancellation List Decoding, 12\,nm FinFET, ASIC Implementation
\end{IEEEkeywords}

\section{Introduction}

In the \gls{eMBB} scenarios of the 5G \gls{NR} standard, polar codes 
\cite{ari_2009} were selected as error-correction codes for the control channels
\cite{3gpp.38.212}. Polar codes achieve the capacity of binary memoryless 
channels under low-complexity \SC decoding at infinite code length. However, 
\SC decoding suffers from a poor error-correction performance in the practical block
length regime. More advanced decoding algorithms, with \SCL decoding
\cite{talvar_2015} being the most prominent one, were developed to overcome 
this limitation at the price of higher decoding complexity and implementation costs. 
However, \SCL decoding approaches \gls{ML} decoding performance with sufficient 
large list size $L$.

Recently, \AED \cite{geielk_2021} for polar codes \cite{geielk_2021a, 
lizha_2021, pilbio_2022} gained attention as a new \gls{ML}-approaching 
algorithm.
In \gls{AED}, an ensemble of $M$ low-complexity decoders (\eg SC) operates on 
different permutations from the code's automorphism group in parallel. The most probable code word is selected as 
output from the $M$ constituent decoders, which improves the overall error-correction performance. 

In \cite{dumsha_2006}, near-\ML performance was achieved for \gls{RM} codes
by running \SCL decoders on permutations of the channel \glspl{LLR} 
corresponding to stage-shuffled factor graphs and combining the different lists 
in each decoding step. This approach was generalized in \cite{doahas_2022}
by initializing the $L$ decoding paths by permutations randomly 
sampled from the full automorphism group of the \gls{RM} code.

The automorphism group of polar codes and its properties were investigated in 
\cite{geielk_2021, geielk_2021a, lizha_2021, pilbio_2022}. 
Through the discovery of suitable automorphisms for polar codes, also more 
advanced polar decoding algorithms, \eg \SCL decoding, can be leveraged.
The new contributions of this work are summarized as follows:
\begin{itemize}
    \item 
        We propose \SCAL decoding of polar codes as a novel method to 
        beneficially utilize permutations selected from the code's
        automorphism group 
        in the \SCL decoding algorithm. 
    \item 
        We analyze the evolution of these permutations during \SCAL decoding and
        provide simulation results to show the capability of channel adaption 
        and the improved error-correction performance, respectively.
    \item 
        We present implementation results of the proposed \SCAL 
        decoder architecture in a \nm{12} FinFET technology and compare them with state-of-the-art, high-throughput \AED and \SCL decoders.
\end{itemize}

The remainder of this paper is structured in four sections: 
Section\,\ref{sec:background} describes the fundamentals of polar codes and the
relevant decoding strategies. The new \SCAL decoding algorithm is presented in 
Section\,\ref{sec:scal}. Results with respect to an analysis of the evolution of
the input permutations, the error-correction performance of \SCAL decoding and
its hardware implementation costs are provided in Section\,\ref{sec:results}.
Section\,\ref{sec:conclusion} concludes this work.

\section{Background}
\label{sec:background}

\subsection{Polar Codes}
\label{subsec:polar}
A polar code $\mathcal{P}(N,K)$ \cite{ari_2009} is a linear block code with code 
length $N=2^n$ and $K$ information bits. The information set~$\mathcal{I}$ defines 
the row indices to obtain the generator matrix~$\bm{G}$ as the rows of
$\bm{G}_N=\bm{G}_2^{\otimes n}$, where $\bm{G}_2 = \bigl[\begin{smallmatrix}
1 & 0 \\
1 & 1 \\
\end{smallmatrix}\bigr]$ denotes the polarization kernel and $\otimes n$ denotes
the $n$-th Kronecker power. The codeword can be obtained from an input 
vector~$\bm{u}$ as ${\bm{x}=\bm{u}\mathbf{G}_N}$, where $\bm{u}$ contains $K$ 
information bits at the positions of~$\mathcal{I}$ and $N-K$ frozen bits (set to
$0$) at the positions of~$\mathcal{F}=\mathcal{I}^C$. 
Polar codes can also be seen as monomial codes. In 
particular, most practical polar codes are decreasing monomial codes, \ie all 
sub-channels in $\mathcal{I}$ obey the partial order according to 
\cite{bardra_2016}, and are thus called decreasing polar codes. They are completely
defined by the minimal information set $\mathcal{I}_\mathrm{min}$ \cite{geielk_2021a}.

\subsection{Successive Cancellation Based Decoding}
\label{subsec:sc}
Polar code decoding can be represented as traversal of a balanced binary tree, 
the \gls{PFT}~\cite{alaksc_2011}. Input to the root node are the channel 
\glspl{LLR}, defined as
${\llr(y_i) = \log(\operatorname{P}(y_i | x_i=0)/\operatorname{P}(y_i | x_i=1))}$, calculated from the 
$N$~received channel values~$\bm{y}$ with $i\in[0,N)$. A node~$v$ in layer
$s$ receives a 
vector~$\bm{\alpha}^v$ of $N_v=2^s$~\glspl{LLR} from its parent node in layer 
$s+1$. The min-sum formulations of $f$- and $g$-functions \cite{lertal_2011} 
are used to obtain the messages passed to the left and the right child nodes, 
$\bm{\alpha}^l$ and $\bm{\alpha}^r$, respectively. The partial-sum 
vector~$\bm{\beta}^v$ is returned to the parent node and calculated by the 
$h$-function combining $\bm{\beta}^l$ and $\bm{\beta}^r$. Thus, \gls{SC}-based 
decoding causes a depth first traversal of the \gls{PFT} with an inherent 
priority to the left child. In the $N$ leaf nodes, $\bm{\beta}^v$ is set to the 
value of the estimated bit~$\hat{u}_i$ as
\begin{equation}
    \hat{u}_i =
    \begin{cases}
        0, &\text{if } i \in \mathcal{F} \text{ or } \alpha^v_i\geq 0 \\
        1, &\text{otherwise.}
    \end{cases}
    \label{eq:u_i}
\end{equation}

\subsection{Successive Cancellation List Decoding}
\label{sec:scl}

In \gls{SCL} decoding \cite{talvar_2015}, lists of
$L$~vectors are passed among the nodes of the \gls{PFT} instead of only passing
vectors~$\bm{\alpha}$ and $\bm{\beta}$. Therefore, ${l\in[0,L)}$ is introduced
to index the $L$ concurrent paths, \eg the \glspl{LLR} of a node~$v$ are
denoted as~$\bm{\alpha}^{v,l}$. For an information bit estimation in layer~$s=0$
of the \gls{PFT}, the decoding path splits, \ie both possible values, $0$ and
$1$, are considered. Consequently, an information bit estimation doubles the
number of decoding paths. Unreliable paths must be rejected by sorting, when
the number of paths exceeds $L$. For this purpose, every bit estimation updates 
a \gls{PM} to rate the reliability of each path. 
These \glspl{PM} are initialized with 0 in \gls{LLR}-based \gls{SCL} decoding
\cite{balpar_2015a}. The \gls{PM} of a path with index~$p \in [0,2L)$ proceeding
input path~$l \in[0,L)$ is updated for the $i$-th bit estimation by
\begin{equation}
    \text{PM}^p_{i} =
    \begin{cases}
        \text{PM}^l_{i-1}
        + \left|\alpha^{v,l}_i \right|,
            &\text{if } \beta^{v,l}_i \neq \text{HDD}(\alpha^{v,l}_i)\\
        \text{PM}^l_{i-1},
            &\text{otherwise,}
    \end{cases}
    \label{eq:pm}
\end{equation}
with \gls{HDD} on~$\bm{\alpha}^v$ defined as
\begin{equation}
    \text{HDD}(\alpha^{v,l}_i) =
    \begin{cases}
        0 &\text{if } \alpha^{v,l}_i \geq 0\\
        1 &\text{otherwise.}
    \end{cases}
    \label{eq:hdd}
\end{equation}
Consequently, the \gls{PM} is a cost function and, thus, the smallest \glspl{PM} 
values belong to the most probable paths which survive the sorting step. The 
most probable path 
is selected as the output of the decoder after the last bit decision.

\subsection{Automorphisms for Polar Code Decoding}
\label{sec:aed}
Automorphisms are permutations that map every code word onto another code word. It 
was shown that the group of affine automorphisms of decreasing polar codes 
equals the \gls{BLTA} group \cite{lizha_2021} defined as 
\begin{equation}
    \bm{z'} = \bm{A}\bm{z} + \bm{b}, 
\end{equation}
where $\bm{A}$ is an invertible, block lower triangular binary matrix, $\bm{b}$ 
an arbitrary binary vector and $\bm{z}, \bm{z'}$ are the binary representations of the bit 
indices before and after permutations, respectively \cite{geielk_2021}. In 
\cite{geielk_2021a}, an algorithm to determine the block profile of $\bm{A}$ for a given~$\mathcal{I}$ is provided. To select proper permutations~$\pi_m$ 
from the automorphism group, a greedy selection method was proposed in 
\cite{kesgei_2023} to pick $M$ permutations from different partitions of 
the automorphisms, \ie the \glspl{EC} \cite{pilbio_2022}.

\section{Successive Cancellation \\Automorphism List Decoding}
\label{sec:scal}

Permuting the received channel data according to automorphisms corresponds to 
different noise realizations. Since there may be permutations that are easier 
to decode, the probability of finding the correct code word is increased in 
\gls{AED} \cite{geielk_2021}. Simultaneously, \AED benefits from the locality 
of the $M$~independent, constituent ensemble decoder cores resulting in 
decreased hardware implementation costs compared to \SCL decoders 
\cite{kesgei_2023}. 

\SCL decoding, on the contrary, splits the decoding paths~$\zeta$ whenever an 
information bit is decoded to overcome error propagation \cite{talvar_2015}. 
In a \emph{Sort \& Select} step, a message exchange takes place to proceed the 
$L$ most probable paths and prevent the exponential increase of the number of 
candidate paths.

To benefit from both approaches of error reduction, \AED with constituent \SCL
decoders is an obvious solution~\cite{geielk_2021}.
However, the messages exchange is then limited to the constituent \SCL decoder
instances and, with one permutation per decoder core, the potential of 
automorphisms is not fully exploited. Furthermore, the selection of the most 
probable candidate in \AED is already inherent in \SCL decoding.
 
In \cite{dumsha_2006, doahas_2022}, near-\ML performance was achieved for 
Reed-Muller codes by running \SCL decoders on permutations of the channel 
\glspl{LLR} and combining the different lists in each decoding step. 
Similarly, \SCL decoding of Polar codes can benefit from the usage of 
automorphisms, which represent different noise realizations of the received 
channel data. For this purpose, an \gls{SCL}-$L$ decoder is initialized with
$L$ permutations~$\pi_l, l \in \left[0,L\right)$ as
\begin{equation}
    \bm{\alpha}^{0,l} = \pi_l\left(\llr(\bm{y})\right),
\end{equation}
instead of only working on the original channel values corresponding to the
identity permutation $\pi_0\left(\llr(\bm{y})\right)$. The permutations are 
sampled from the automorphism \glspl{EC} of the code by a greedy selection 
method as described in \cite{kesgei_2023}.
Consequently, the list of candidates is filled from the start of \SCL 
decoding and does not only start expanding with path splitting in the
first nodes containing information bits. 

While the \PMs of all candidates (\ie decoding paths $\zeta_l$) are already 
updated in the left-most frozen nodes, path splitting is superposed in 
non-frozen nodes. Thus, the different noise realizations of the 
permutations compete against each other whenever the expanded list of 
candidates is sorted and pruned to $L$ in the sorters of the non-frozen nodes.
This message exchange in the \emph{Sort \& Select} steps provides a gain 
compared to the independent decoding of the permutations in \gls{AED}. Thus, 
\SCAL has the ability to adapt to the channel conditions automatically, since
either the candidates generated by path splitting or stemming from the 
permutations persist preferably.

Finally, the \SCAL decoder outputs the most reliable candidate, \ie the path 
with the smallest \gls{PM}, which is found as the first entry (index $0$) in 
the final sorted list of candidates. Depending on its origin $o(\bm{\hat{x}_0})$, the reverse permutation $\pi^{-1}_{o(\bm{\hat{x}_0})}$ of 
the corresponding input permutation is applied as
\begin{equation}
    \bm{\hat{x}}=\pi^{-1}_{o(\bm{\hat{x}_0})}\left(\bm{\beta}^{0,0}\right)
\end{equation}
to obtain the decoded code word.

\begin{figure}
    \centering
    \vspace{-1pt}
    \includegraphics[width=0.70\linewidth]
        {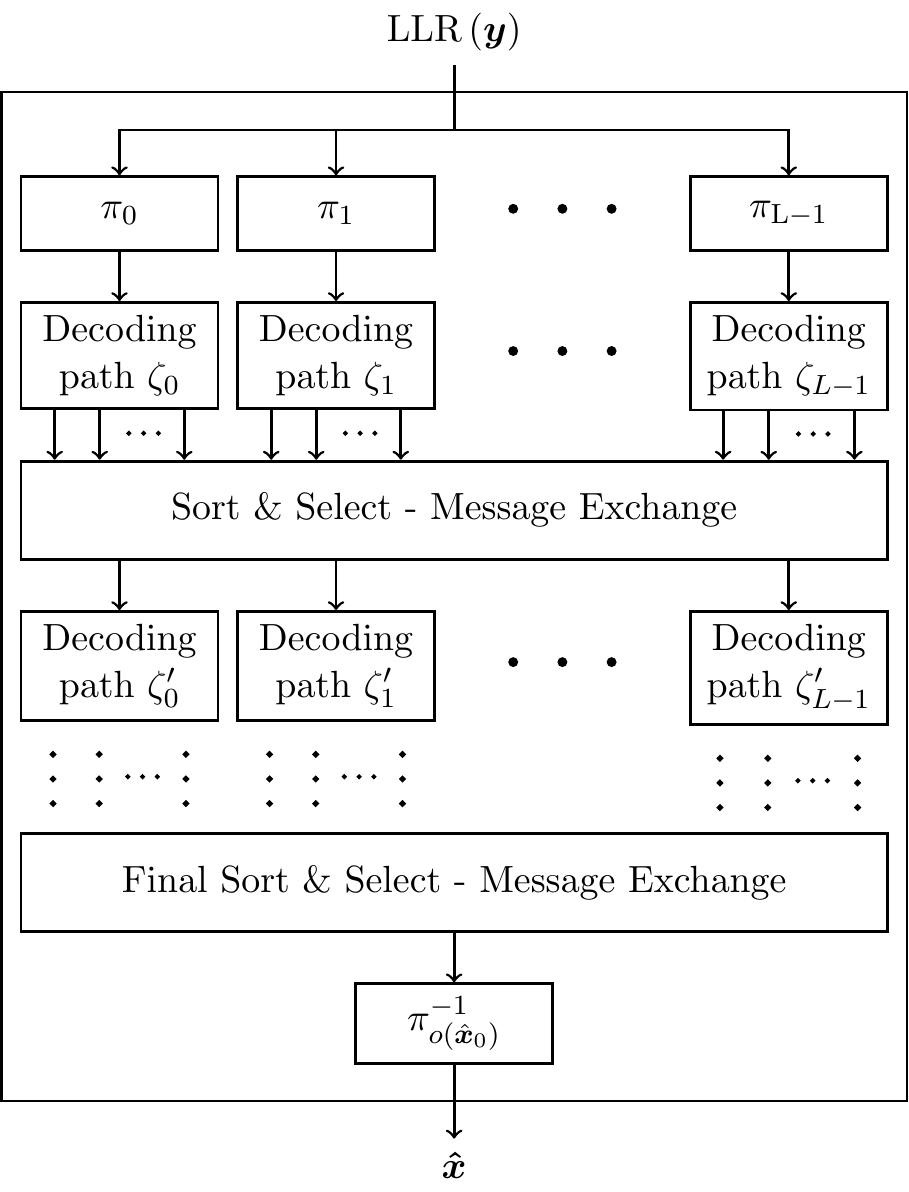}
    \vspace{-6pt}
    \caption{\SCAL decoding architecture}
    \label{fig:scal_architecture}
    \vspace{-8pt}
\end{figure}

The architecture of an \SCAL decoder is shown in 
Fig.\,\ref{fig:scal_architecture} indicating the parallel decoding of paths 
$\zeta_l$ until leaf nodes split the decoding paths. The decoding of each path 
$\zeta$ produces multiple candidates. Based on the corresponding \glspl{PM}, 
the \emph{Sort \& Select} step consolidates the $L$ most reliable paths by the 
message exchange of the incoming candidate paths. 
It is noteworthy, that \SCAL decoding can adopt the known, state-of-the-art 
optimizations for complexity reduction of \SCL decoding without 
error-correction performance degradation.

\section{Results}
\label{sec:results}

The results presented in this section are based on the polar code
$\mathcal{P}(128,60)$ with $\mathcal{I}_\mathrm{min}=\{27\}$, which exhibits a
block profile $\bm{s}=(3,4)$ of its \gls{BLTA} group and is therefore well
suited for \AED \cite{kesgei_2023} and \SCAL decoding. The permutation 
selection follows \cite{kesgei_2023} to pick one automorphism from $L$ 
\glspl{EC}. Since the same code was used in \cite{kesgei_2023}, this selection
enables a direct comparison to \gls{AED}.

The Fast Simplified \SCAL (Fast-S\gls{SCAL}) and \gls{FSSCL} decoders use 
optimized nodes according to \cite{johkes_2022, johkes_2022a} and a 
quantization of \SI{6}{\bit} for \glspl{LLR} and \SI{8}{\bit} for \glspl{PM}, 
for both, \FER simulations and hardware implementations. 

Our simulation model applies \gls{BPSK} modulation and uses an \gls{AWGN} 
channel. Up to $10^8$\,pseudo-random blocks with a limit of $10^3$\,erroneous
decoded block are simulated.

\subsection{Analysis of Automorphism Evolution}

\begin{figure}
    \centering
    \includegraphics[width=1\linewidth]
        {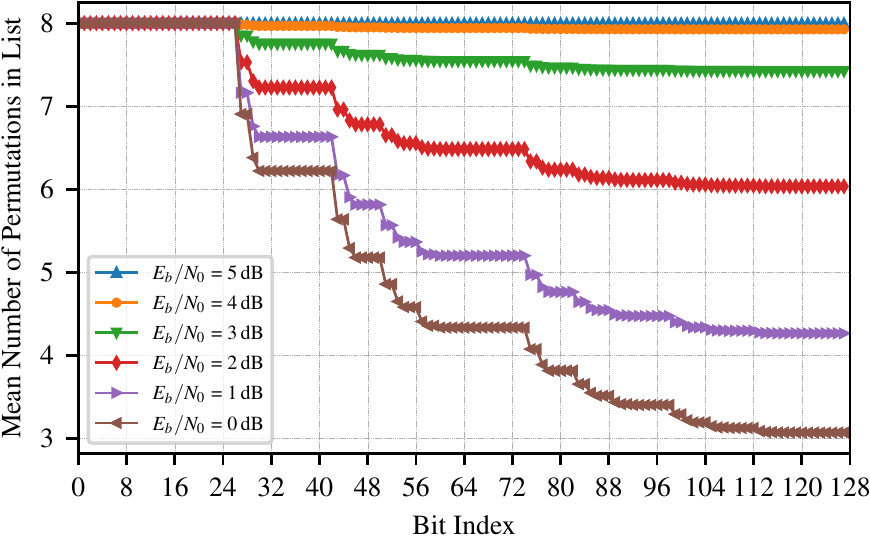}
    \vspace{-16pt}
    \caption{
        Evolution of the number of different permutations $\pi_l$ vs. the bit 
        indices during \gls{SCAL}-$8$ decoding of $\mathcal{P}(128,60)$}
    \label{fig:plot_perm_num_scal}
    \vspace{-10pt}
\end{figure}

To evaluate the novel \SCAL decoding algorithm, the evolution of the 
permutations is analyzed. The permutations can be seen as different noise 
realizations, and the ones which are easier to decode prevail. Thus, the mean 
number of different permutations within the list of surviving candidates 
decreases during \SCAL decoding. This effect is shown in 
Fig.\,\ref{fig:plot_perm_num_scal} for the $\mathcal{P}(128,60)$ under 
\gls{SCAL}-$8$ decoding at different \SNR points. 
The figure also shows that the permutations dominate 
classical candidate generation in \SCL decoding with increasing \gls{SNR}.

\begin{figure}
    \centering
    \vspace{3pt}
    \resizebox{\linewidth}{!}{\input{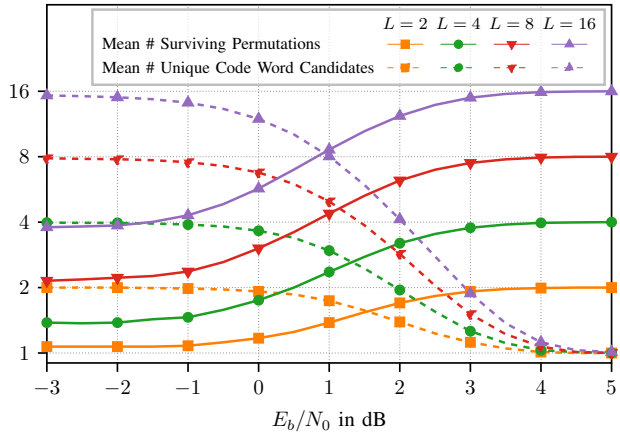}}
    \vspace{-17pt}
    \caption{Average number of unique final code word candidates and average number of permutations vs. \gls{SNR} for \gls{SCAL} decoding of $\mathcal{P}(128,60)$}
    \label{fig:perm_snr_scal}
    \vspace{-2pt}
\end{figure}

Unlike in \SCL decoding, the final code word candidates (after undoing the permutation) are not necessarily different in \SCAL decoding.
Fig.\,\ref{fig:perm_snr_scal} shows the average number of unique code word candidates and the number of different permutations of the surviving paths before the final \emph{Sort \& Select} stage.
For low \gls{SNR}, a large list diversity can be observed, as the number of unique code word candidates is almost identical to the maximum number of paths $L$. 
However, these candidates stem from only a small number of permutations, indicating that the \gls{SCL} branching dominates candidate generation in this regime.
Conversely, at high \gls{SNR}, the decoder only rarely branches and, thus, behaves more like \AED where almost all $L$ initial permutations survive until the final selection.
However, this results in a low list diversity, as all decoding paths converge to the same (correct) code word candidate most of the time.

\subsection{Error-Correction Performance}
\label{subsec:fer}

\begin{figure}
    \centering
    \includegraphics[width=1\linewidth]{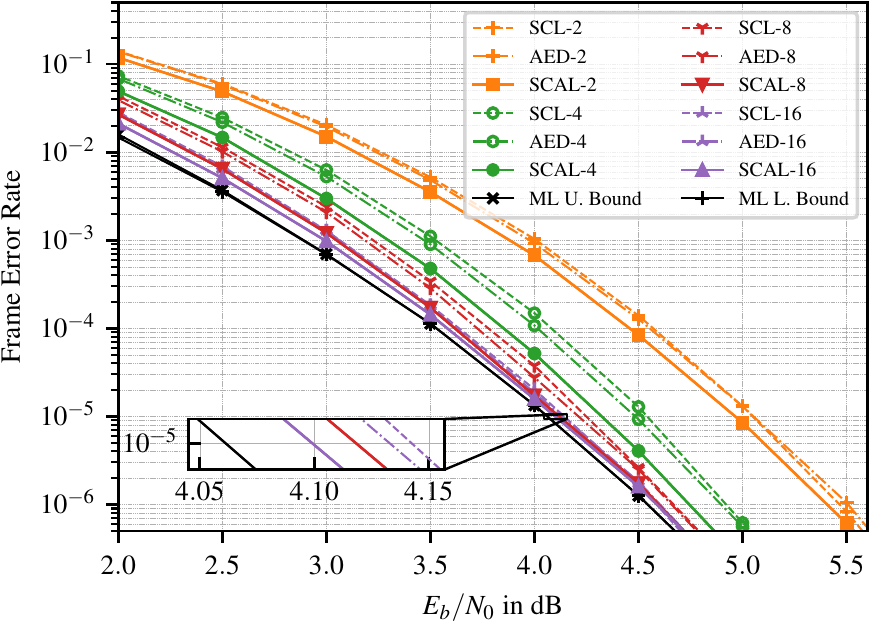}
    \vspace{-14pt}
    \caption{\FER vs. \SNR for \gls{SCL}, \AED and \SCAL decoding of
        $\mathcal{P}(128,60)$
    }
    \label{fig:plot_fer_scal}
    \vspace{-12pt}
\end{figure}

The plot in Fig.\,\ref{fig:plot_fer_scal} shows the \glspl{FER} of \SCAL 
decoding, \SCL decoding and \AED for $L/M \in \{2,4,8,16\}$. The upper and 
lower bounds for \gls{ML} decoding performance are also given, which are
observed by performing SCL decoding with $L = 128$. Whenever the decoded 
codeword is closer to the received channel values than the correct codeword, an 
\gls{ML} decoder would fail. Thus, counting these cases provides the upper 
\gls{ML} bound. Similarly, the lower bound is derived with the additional 
condition that the final list of candidates contains the correct codeword.
In Fig.\,\ref{fig:plot_fer_scal}, both bounds coincide.

It can be seen that \gls{SCAL}-$2$ provides a gain of \dB{0.1} over \AED and 
\SCL decoding.
At an FER of $10^{-5}$, SCAL\=/$4$ has a performance gain of \dB{0.22} and 
\dB{0.16} compared to SCL-$4$ and AED-$4$ and a gap of \dB{0.074} and
\dB{0.11} to SCL-$8$ and AED-$8$, respectively.
At the same FER, SCAL-$8$ outperforms SCL-$16$ and AED-$16$ by \dB{0.024} and
\dB{0.015}, respectively, and the gap to the \gls{ML} bound is \dB{0.06}.
To summarize, \SCAL decoding outperforms the error-correction performance of 
\AED and \SCL decoding significantly for all list sizes/degrees of parallelism.

\subsection{Hardware Implementation}

In addition to the improved error-correction performance, also the 
implementation costs of the proposed decoders have to be considered. 
Thus, we implemented different \gls{AED}, \SCL and \SCAL decoders.
All decoder implementations are unrolled and fully pipelined architectures for 
a target frequency of \SI{500}{\mega\hertz}. Synthesis was executed with
\textit{Design Compiler} and \gls{PAR} with \textit{IC-Compiler}, both from
\textit{Synopsys}, in a \nm{12} FinFET technology from Global Foundries under 
worst case \gls{PVT} conditions (\SI{125}{\degreeCelsius}, \SI{0.72}{V}) for
timing and nominal case \gls{PVT} (\SI{25}{\degreeCelsius}, \SI{0.8}{V}) for 
power. Power values stem from post-\gls{PAR} netlist simulations with 
back-annotated wiring data and test data at $E_b/N_0=\text{\dB{4}}$.

In fully pipelined architectures, the coded throughput is calculated as
$T_c=f\cdot N$, with clock frequency $f$. Metrics for comparison are area 
efficiency $\mu_A=T_c/A$, energy efficiency $\mu_E=P/T_c$, with $A$ being the 
area and $P$ the power of the implementation, and power density 
$P/A=\mu_E\cdot\mu_A$.

\begin{table}[t]
    \centering
    \caption{Implementation results of \SCAL and AE decoders
    }
    \label{tab:12nm_scal_ae}
    \begin{tabular}{lcccc}
\toprule
                          &\textbf{SCAL} &AED   &\textbf{SCAL} &AED\\
$M/L$                     &\textbf{4}    &8     &\textbf{8}    &16\\
\midrule
Frequency {[}MHz{]}       &498           &498   &500           &495 \\
Throughput {[}Gbps{]}     &63.8          &63.7  &64.0          &63.3 \\
Latency {[}ns{]}          &48.2          &22.1  &64.0          &22.2 \\
Area {[}mm$^2${]}         &0.128         &0.170 &0.381         &0.338 \\
\textbf{Area Eff.}%
\textbf{{[}Gbps/mm$^2${]}}
                          &\textbf{496.5 }
                          &\textbf{375.1 }
                          &\textbf{168.2 }
                          &\textbf{187.0 }\\
Utilization {[}\%{]}      &72            &74    &72            &73 \\
Power Total {[}W{]}       &0.194         &0.320 &0.552         &0.640 \\
\textbf{Energy Eff.}%
\textbf{{[}pJ/bit{]}}
                          &\textbf{3.04 }
                          &\textbf{5.03 }
                          &\textbf{8.62 }
                          &\textbf{10.12 }\\
Power Density %
{[}W/mm$^2${]}            &1.51          &1.89  &1.45          &1.89 \\
\bottomrule
\end{tabular}

    \vspace{-8pt}
\end{table}

\subsubsection{SCAL decoding vs. AED}\hfill\\
The \gls{PAR} results for \SCAL decoder and \AED implementations are presented 
in Table\,\ref{tab:12nm_scal_ae}. For a fair comparison of the implementation 
costs of \SCAL and AE decoders, the architectures having comparable 
error-correction performance are selected. Thus, \gls{SCAL}-$4$ is compared 
with \gls{AED}-$8$, and \gls{SCAL}-$8$ with \gls{AED}-$16$. It can be seen, 
that the \SCAL decoders have a better energy efficiency than their \AED 
counterparts.
The area efficiency of \gls{SCAL}-$4$ is $1.32\times$ better than 
$\mu_\text{A}$ of \gls{AED}-$8$. Comparing $\mu_\text{A}$ of \gls{SCAL}-$8$ and
\gls{AED}-$16$, the \SCAL decoder is $0.90\times$ worse.
All presented \SCAL decoders have a better power density than \gls{AED}. In 
terms of latency, \AED is always at an advantage, since its latency is defined 
by the latency of the \SC decoder cores. 

\subsubsection{SCAL decoding vs. SCL decoding}\hfill\\
\begin{figure}
    \vspace{-8pt}
    \centering
    \subfloat[SCL: \mmsq{0.267}]{%
        \includegraphics[height=0.44\columnwidth]
            {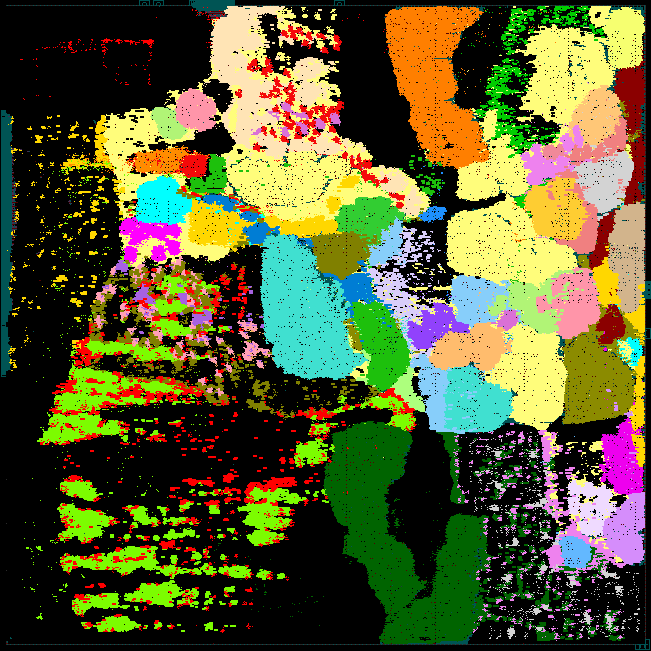}
        \label{fig:layout_scl}}
    \hfill
    \centering
    \subfloat[SCAL: \mmsq{0.381}]{%
        \includegraphics[height=0.53\columnwidth]
            {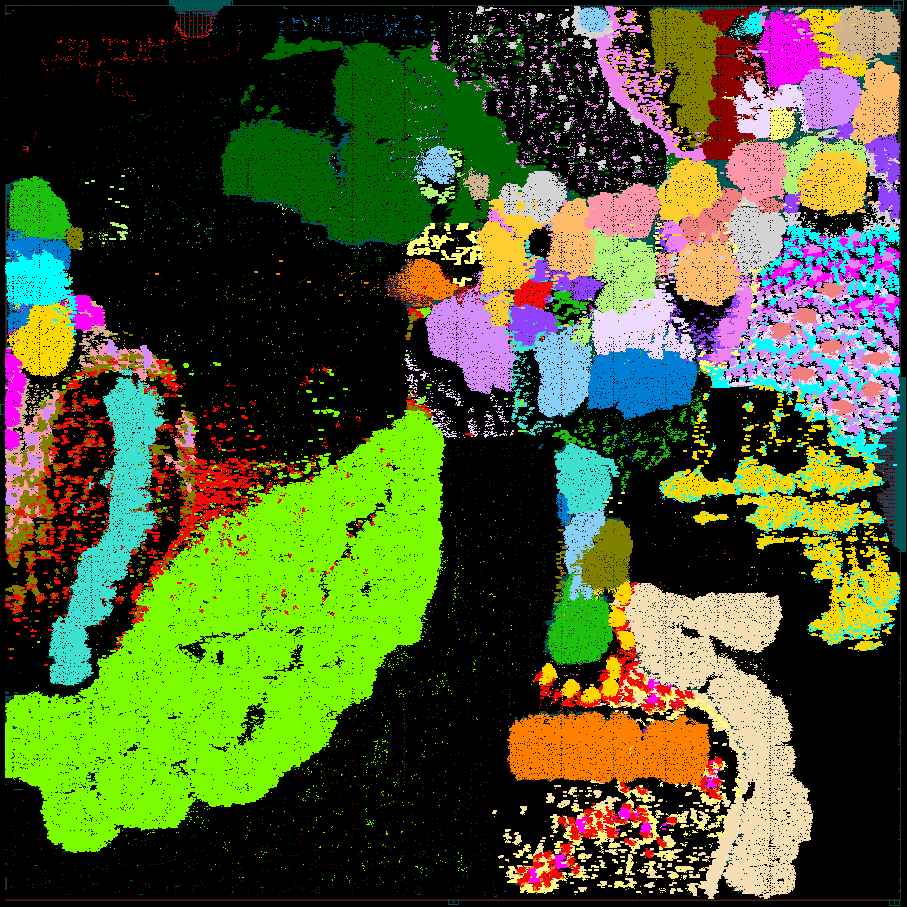}
        \label{fig:layout_scal}}
    \caption{
        Layouts of \gls{SCL}-$8$ decoder \cite{johkes_2022a} (a) and
        \gls{SCAL}-$8$ decoder (b), both for $\mathcal{P}(128,60)$;
        different colors represent computational kernels, delay line memory
        is colored in black; the layouts have the same scaling.
    }
    \label{fig:layout_scal_scl_8}
    \vspace{-6pt}
\end{figure}
The comparison of \SCAL and \SCL decoder implementations aims for showing the 
overhead caused by generating and processing $L$ permutations of the input 
\glspl{LLR} in the \SCAL decoder. Thus, two decoders with equal $L$ are 
compared, respectively. The threshold parameters for the optimized \SPC nodes 
\cite{johkes_2022a} are set to preserve the error-correction performance of each
decoder, \ie $S_\text{SPC}=\{2,2\}$ and $k_\text{SPC}=\{2,3\}$ for 
\gls{SCAL}-$\{4,8\}$, and $S_\text{SPC}=\{3,4\}$ and $k_\text{SPC}=\{2,3\}$ for 
\gls{SCL}-$\{4,8\}$, respectively. The corresponding results are given in 
Table\,\ref{tab:12nm_scal_scl}.

Fig.\,\ref{fig:layout_scal_scl_8} shows the two layouts of the \gls{SCL}-$8$
decoder\,(\ref{fig:layout_scl}) and the \gls{SCAL}-$8$ 
decoder\,(\ref{fig:layout_scal}) with equal area scaling. Whenever possible, equal coloring is used for the computational kernels in
both implementations. The most conspicuous difference is the size of the cells
highlighted in lime-green, which correspond to the first $g$-function of the 
decoder in the root node. Since in the \SCL decoder, only one input vector 
needs to be processed (in $L$ variants), in \SCAL decoding the input to this 
$g$-function is a list of $L$~\LLR vectors. The corresponding delay-line memory 
(colored in black for all stages) is also $\times L$ greater. Thus, costs in
area efficiencies are $\mu_\text{A, SCAL-$4$}\approx\mu_\text{A, SCL-$4$}
\times0.78$ and $\mu_\text{A, SCAL-$8$} \approx \mu_\text{A, SCL-$8$}
\times0.70$, and in energy efficiencies $\mu_\text{E, SCAL-$4$} \approx 
\mu_\text{E, SCL-$4$}\times1.21$ and $\mu_\text{E, SCAL-$8$} \approx 
\mu_\text{E, SCL-$8$}\times1.27$, respectively.
However, the usage of the permutations brings a gain in the error-correction 
performance (section\,\ref{subsec:fer}) for which reason \SCAL decoders 
outperform comparable \SCL decoders.

\begin{table}[t]
    \centering
    \caption{Implementation results of \SCAL and \SCL decoders
    }
    \vspace{-2pt}
    \label{tab:12nm_scal_scl}
    \begin{tabular}{lcccc}
\toprule
                          &SCL   &\textbf{SCAL}&SCL   &\textbf{SCAL}\\
$L$                       &4     &\textbf{4}   &8     &\textbf{8}\\
\midrule
Frequency {[}MHz{]}       &500   &498          &500   &500 \\
Throughput {[}Gbps{]}     &64.0  &63.8         &64.0  &64.0 \\
Latency {[}ns{]}          &44.0  &48.2         &62.0  &64.0 \\
Area {[}mm$^2${]}         &0.101 &0.128        &0.267 &0.381 \\
\textbf{Area Eff. {[}Gbps/mm$^2${]}}
                          &\textbf{632.8 }
                          &\textbf{496.5 }
                          &\textbf{239.5 }
                          &\textbf{168.2 }\\
Utilization {[}\%{]}      &73    &72           &73    &72 \\
Power Total {[}W{]}       &0.161 &0.194        &0.434 &0.552 \\
\textbf{Energy Eff. {[}pJ/bit{]}}
                          &\textbf{2.52 }
                          &\textbf{3.04 }
                          &\textbf{6.79 }
                          &\textbf{8.62 }\\
Power Density {[}W/mm$^2${]}
                          &1.60  &1.51         &1.63  &1.45 \\
\bottomrule
\end{tabular}

    \vspace{-8pt}
\end{table}

\section{Conclusion}
\label{sec:conclusion}

In this paper, \SCAL decoding is presented as a novel method to benefit from
using automorphisms of polar codes together with \SCL decoding. The proposed 
\SCAL decoding algorithm outperforms \AED and \SCL decoding with respect to the 
error-correction performance. Regarding the hardware implementation costs, \SCAL 
decoders benefit from the possible reduction of the list size and therefore 
compete with \AED and \SCL decoder implementations with comparable error 
correction.

\bibliographystyle{IEEEtran}
\bibliography{main}

\end{document}